\documentclass[a4paper,10pt]{article}
\usepackage[utf8x]{inputenc}
\usepackage{graphicx}
\usepackage{amsmath}
\usepackage{cite}
\usepackage{authblk}
\usepackage{color}
\usepackage[margin=3cm]{geometry}

\title{\bf Climate network and complexity based El~Ni\~no forecast for 2022}
\date{}

\author[1] {Josef Ludescher}
\author[2,1] {Jun Meng}
\author[3,1] {Jingfang Fan}

\tiny
\affil[1] {Potsdam Institute for Climate Impact Research, 14412 Potsdam, Germany}
\affil[2] {School of Science, Beijing University of Posts and Telecommunications, Beijing 100876, China}
\affil[3] {School of Systems Science, Beijing Normal University, 1000875 Beijing, China}

\begin{document}

\maketitle

\begin{abstract}
The El~Ni\~no Southern Oscillation (ENSO) is the most important driver of interannual global climate variability and can trigger extreme weather events and disasters in various parts of the globe. Recently, we have developed two approaches for the early forecasting of El~Ni\~no. The climate network-based approach allows forecasting the onset of an El~Ni\~no event about 1 year ahead \cite{Ludescher2013}. 
The complexity-based approach allows additionally to forecast the magnitude of an upcoming El~Ni\~no event in the calendar year before \cite{Meng2019}. Here we communicate the forecasts of both methods for 2022. 
\end{abstract}

\section{The El~Ni\~no Southern Oscillation}

The El~Ni\~no-Southern Oscillation (ENSO) phenomenon \cite{Clarke08, Sarachik10, Power2013, Dijkstra2005, Wang2017,Timmermann2018, McPhadden2020} 
can be perceived as a self-organized dynamical see-saw pattern in the Pacific ocean-atmosphere system, featured by rather irregular warm (``El~Ni\~no'') and cold (``La Ni\~na'') excursions from the long-term mean state. The ENSO phenomenon is quantified by the  Oceanic Ni\~no Index (ONI), which is based on the average sea-surface temperature (SST) in the Ni\~no3.4 region in the Central Pacific (see Fig. 1). 

The ONI is defined as the three-month running-mean SST anomaly in the Ni\~no3.4 region and is a principal measure for monitoring, assessing and predicting ENSO. We will refer to the ONI also as the Ni\~no3.4 index. An El~Ni\~no episode is said to occur when the index is at least 0.5°C above the climatological average for at least 5 months. A regularly updated table of the ONI can be found at \cite{NOAA}.

\begin{figure}[]
\begin{center}
\includegraphics[width=9cm]{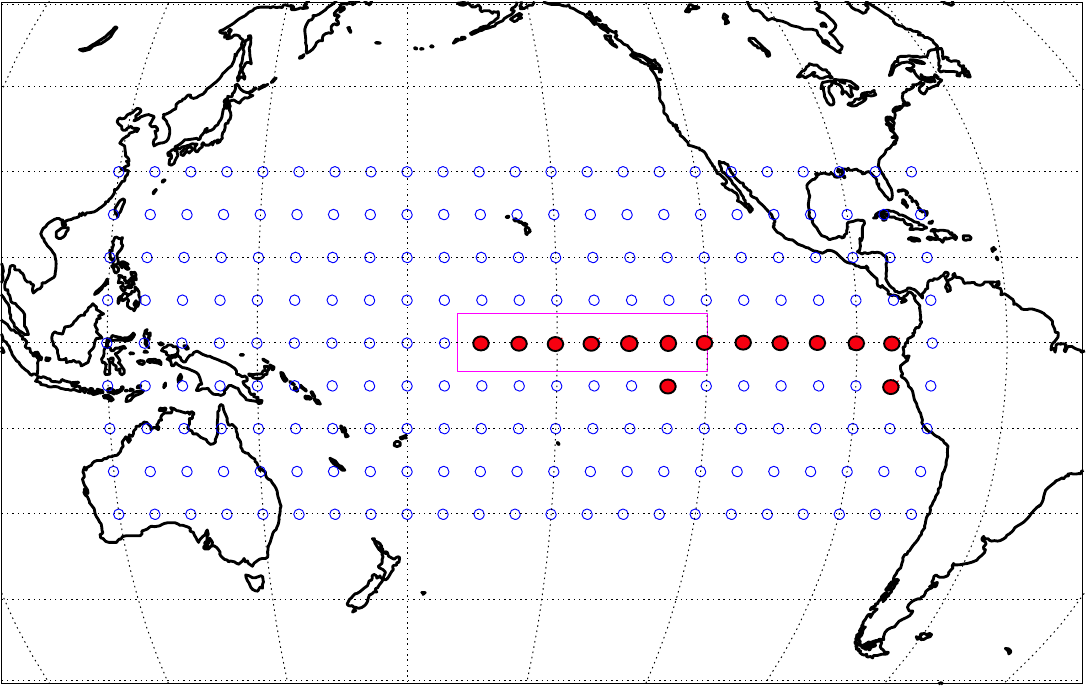}
\caption{The  ONI and the ``climate network''. The network consists of 14 grid
points  in the ``El~Ni\~no basin'' (solid red symbols) and 193 grid points outside this
domain (open symbols). The red
rectangle denotes the area where the ONI (Ni\~no3.4 index) is measured. The grid points are considered as the nodes of the climate network that we use here to forecast El~Ni\~no events. Each node inside the El~Ni\~no basin is linked to each node outside the basin. The nodes are
characterized by their surface air temperature (SAT), and the link strength between the nodes is determined from their cross-correlation (see below). Figure from \cite{Ludescher2013}.}
\label{fig1}
\end{center}
\end{figure}

Since strong El~Ni\~no episodes can wreak havoc in various parts of the world (through 
extreme weather events and other environmental perturbations)  \cite{Wen2002,Corral10,Donnelly07,Kovats03,Davis2001,McPhadden2020},
early-warning schemes based on robust scientific evidence are highly desirable.
Sophisticated global climate models taking into account the atmosphere-ocean coupling, as well as statistical approaches like the dynamical systems schemes approach, autoregressive models and pattern-recognition techniques, have been proposed to forecast the pertinent index with lead times between 1 and 24 months 
\cite{Clarke08,Cane86,Latif94,Tziperman97,Kirtman98,Landsea00,Kirtman03,Fedorov03,Muller04,Chen04,Palmer06,Luo08,Yeh09,Chekroun11,Galanti03,Chen08,Penland1995,Chapman2015, Nootboom2018, Meng2018, Ham2019}. 

Unfortunately, so far, the forecasting methods in operation have quite limited anticipation power. In particular, they generally fail to overcome the so-called ``spring barrier'' (see, e.g., \cite{Webster1995,Goddard2001}), which shortens their warning time to around 6 months.

To resolve this problem, we have recently introduced two alternative forecasting approaches \cite{Ludescher2013,Meng2019}, which considerably extend the probabilistic prediction horizon. The first approach \cite{Ludescher2013} (see also \cite{Ludescher2019,Ludescher2021a}) is  based on complex-networks analysis \cite{Tsonis2006,Yamasaki2008,Donges2009,Gozolchiani2011,Dijkstra2019,Fan2020,Ludescher2021,Fan2022}. The method provides forecasts for the onset of an El~Ni\~no event, but not for its magnitude, in the year before the event starts. The second approach \cite{Meng2019} 
relies on the System Sample Entropy (SysSampEn), i.e., an information entropy, in the Ni\~no3.4 area. It provides forecasts for the onset and magnitude of an El Ni\~no event at the end of the foregoing year. 

Here we communicate the forecasts of both methods for 2022. The network-based method predicts the absence of an El~Ni\~no event in 2022 with 90\% probability. The complexity-based method predicts the onset of an El~Ni\~no event with $1.90\pm 0.23$°C magnitude in 2022 with 69\% probability. 
There were 6 analog cases in the forecasted and hindcasted past (1984-2021), where an El~Ni\~no onset prediction of the complexity-based method was not matched by the network-based method. In 2 of these cases, an El Ni\~no did start (2004, 2006) and in 4 cases, it did not start (1985, 1993, 2000, 2012). 
Based on this, it appears more likely, that an El~Ni\~no will not start in 2022. However, if an El~Ni\~no should start in 2022, the complexity-based method forecasts it to have a large magnitude.

\section{Climate network-based forecasting}

\subsection{The network-based forecasting algorithm}

The climate network-based approach exploits the remarkable observation that a large-scale cooperative mode linking the ``El~Ni\~no basin'' (i.e., the equatorial Pacific corridor) and the rest of the Pacific ocean (see Fig. 1) builds up in the calendar year before an El~Ni\~no event. In the description and discussion of the method we follow closely \cite{Ludescher2021a}. An appropriate measure for the emerging cooperativity can be derived from the time evolution of the teleconnections (``links``) between the atmospheric temperatures at the grid points (''nodes``) inside and outside of the El~Ni\~no basin. The strengths of those links are represented by the values of the respective cross-correlations (for details, see \cite{Ludescher2013,Ludescher2019}).
The crucial entity is the mean link strength $S(t)$ as obtained by averaging over all individual links in the network at a given instant $t$ \cite{Ludescher2013,Ludescher2019}. $S(t)$ rises when the cooperative mode builds up and drops again when this mode collapses rather conspicuously with the onset of the El~Ni\~no event. The rise of $S(t)$ in the year before an El~Ni\~no event starts serves as a precursor for the event. 

For the sake of concrete forecasting, we employed in \cite{Ludescher2013} daily surface air temperature (SAT) anomalies for the 1950-2011 period. The data have been obtained from the National Centers for Environmental Prediction/National Center for Atmospheric Research Reanalysis I project \cite{reanalyis1,reanalyis2}.
The optimized algorithm \cite{Ludescher2013,Ludescher2019} involves an empirical decision threshold $\Theta$.  Whenever $S$ crosses $\Theta$ from below while the most recent ONI is below 0.5°C, 
the algorithm sounds an alarm and predicts an El~Ni\~no inception in the following year. 
For obtaining and testing the appropriate thresholds, we divided the data into two halves. In the first part (1950-1980), which represents the learning phase, all thresholds above the temporal mean of $S(t)$ were considered and the optimal ones, i.e., those that lead to the best predictions in the learning phase, have been determined. 
We found that $\Theta$-values between $2.815$ and $2.834$ lead to the best performance \cite{Ludescher2013}, with a false alarm rate of 1/20. In the second part of the data set (1981-2011), which represents the prediction (hindcasting) phase, the performance of these thresholds has been tested. We found that the thresholds between 2.815 and 2.826 gave the best results (see Fig. 2, where $\Theta=2.82$). The alarms were correct in 75\% and the non-alarms in 86.4\% of the cases. For $\Theta$-values between $2.827$ and $2.834$, the performance was only slightly weaker. 
We like to note that for all calculations in the prediction phase, e.g., of the climatological average, only data from the past up to the prediction date have been considered.

\begin{figure}[]
\begin{center}
\includegraphics[width=14cm]{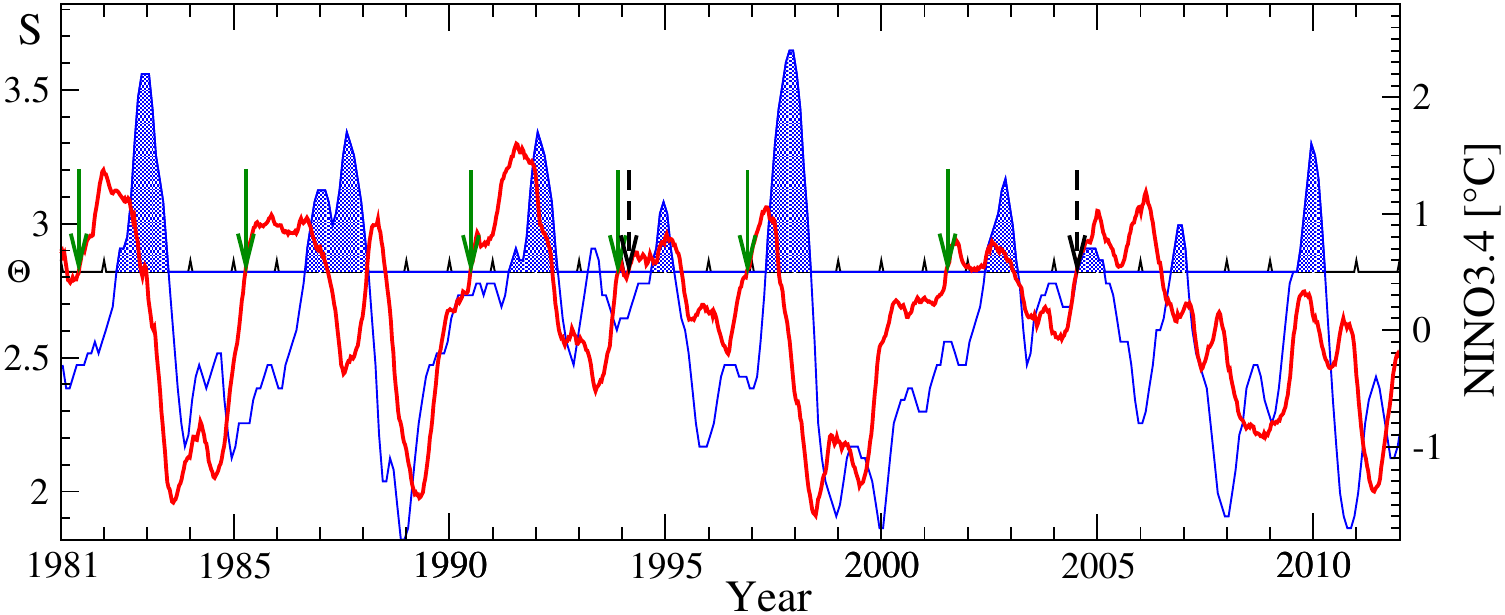}
\caption{The network-based forecasting scheme (hindcasting phase).  We compare the average link strength $S(t)$   
in the climate network (red curve) with a decision threshold $\Theta$ (horizontal line, here $\Theta = 2.82$), (left scale), and the standard Ni\~no3.4 index (ONI), (right scale), between January 1981 and December 2011. 
When the link strength crosses the threshold from below, and the last available ONI is below 0.5°C,
we give an alarm and predict that an El~Ni\~no episode will start in the following calendar year. 
The El~Ni\~no episodes (when the Ni\~no3.4 index is at or above $0.5°$C for at least 5 months) are shown by the solid blue areas. 
Correct predictions are marked by green arrows and false alarms by dashed arrows. 
Between 1981 and 2011, there were 9 El Ni\~no events. The algorithm generated 8 alarms, and 6 were correct. 
In the whole period between 1981 and December 2021, there were 11 El Ni\~no events. The algorithm generated 11 alarms and 8 of these were correct.}
\label{fig2}
\end{center}
\end{figure}

\subsection{Forecasting the next El~Ni\~no (2011 - present)}

Based on this hindcasting capacity, the approach has already been used in \cite{Ludescher2014}
to extend the prediction phase from the end of 2011 until November 2013 and later in \cite{Ludescher2019} and \cite{Ludescher2021a} until December 2020.  
We like to emphasize that in the forecasting phase, the algorithm does not contain any fit parameters since the decision thresholds are fixed and the mean link strengths only depend on the atmospheric temperature data.

Nine out of ten real predictions into the future for the years 2012-2021 turned out to be correct (see Fig. 3). 
These predictions were not trivial. For example, as late as August 2012, the Climate Prediction Center/International Research Institute for Climate and Society (CPC/IRI) Consensus Probabilistic ENSO forecast yielded a 3 in 4 likelihood for an El~Ni\~no event in 2012, which turned out to be incorrect only a few months later \cite{IRI,NOAA}. In contrast, the network approach already forecasted the absence of an El~Ni\~no at the end of 2011.  
In 2013, our algorithm predicted the return of an El~Ni\~no event in 2014, since, in
September 2013,  $S(t)$ transgressed the alarm threshold band while the last available ONI (JJA 2013) was below 0.5°C, indicating the return of El~Ni\~no in 2014 (see Fig. 3). This early prediction was also correct: The El~Ni\~no event started in November 2014 (and ended in May 2016) \cite{NOAA}. For comparison, the furthest into the future (ASO 2014) IRI/CPC plume forecast probabilities in December 2013 were 46\% for a neutral event, 44\% for an El~Ni\~no, and 10\% for a La Ni\~na.
In 2014, 2015, 2016 and 2018, $S(t)$ did not cross the threshold from below, thus indicating the absence of an El~Ni\~no onset in the respectively following years, which all turned out to be correct (see Fig. 3). In November 2017, $S(t)$ transgressed from below the lower threshold band between $S=2.815$ and 2.826. Since the last ONI, for ASO 2017, was below 0.5°C (-0.4°C), this indicated the return of El~Ni\~no in 2018 (see Fig. 3), which turned out to be correct.

\begin{figure}[]
\begin{center}
\includegraphics[width=11cm]{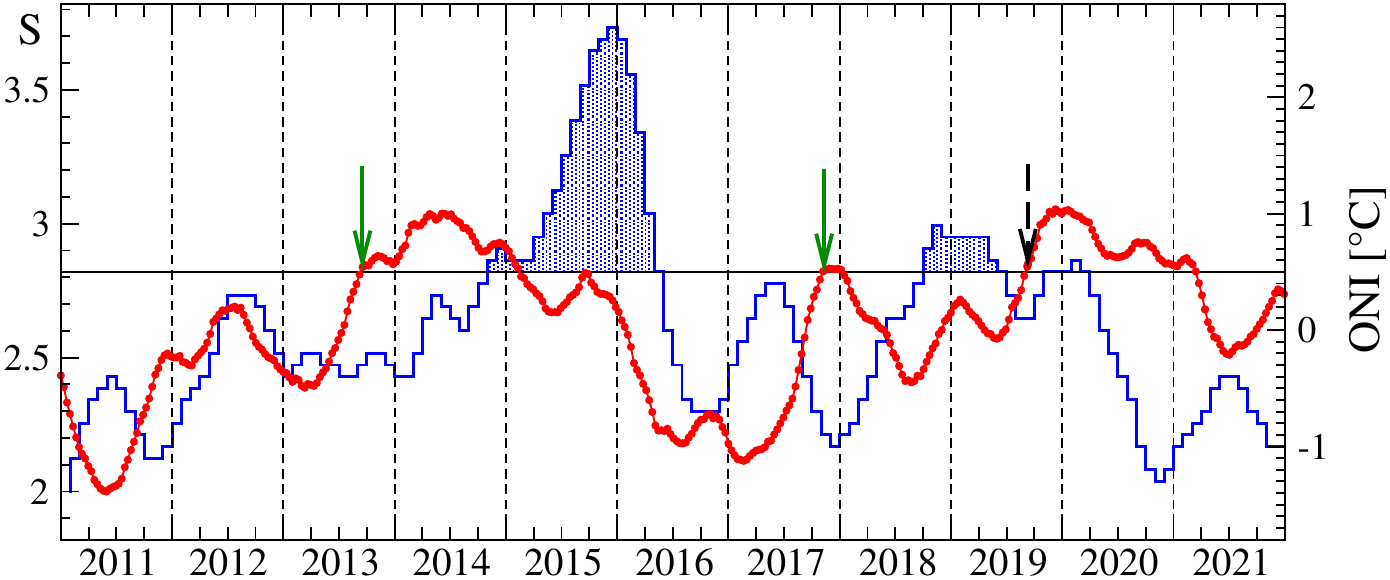}
\caption{ 
The climate network-based forecasting phase. Same as Fig. 2 but for the 
period between January 2011 and December 2021. In 2021 the average link strength $S(t)$ did not cross the threshold from below, i.e., no alarm for an El~Ni\~no in 2022 was given. 
In the hindcasting and forecasting phase (1981-2020), our algorithm predicted 29 times (26 of which were correct) the absence of an El~Ni\~no onset. Thus the likelihood based on the past performance of the climate network approach for the absence of an El~Ni\~no in 2022 is 90\%.
}
\label{fig3}
\end{center}
\end{figure}

In September 2019, $S(t)$ transgressed all thresholds, while the last ONI, JJA 2019, was below 0.5°C (0.3°C). This indicated the onset of an El~Ni\~no in 2020 with 80\% probability. This prediction turned out to be incorrect since in 2020 a La Ni\~na started. This was the first forecast error of the method and the first hindcast of forecast error since 2009 when the method missed the 2009/10 El~Ni\~no.

In 2020 $S(t)$ stayed above the threshold throughout the year (see Fig. 3), thus predicting the absence of an El~Ni\~no in 2021 with 89\% probability \cite{Ludescher2021a}. This prediction turned out to be correct since a second La Ni\~na started in 2021 (see Fig. 3).

Before coming to the forecast for the next year, let us discuss the probability that the same or a better outcome could be obtained by simply guessing using the climatological El Ni\~no probabilities. In the 72 years between 1950 and 2021, 23 El Ni\~nos have started. Accordingly, the probability that an El Ni\~no starts in a certain year is $23/72$. The probability to correctly forecast the El Ni\~no onsets or their absences between 2012 and 2021 (the forecasting phase) is, therefore,
$p=(23/72)^2 (49/72)^8\approx 0.00470$ and having 9 out of 10 correct forecasts with one false alarm $p=8(23/72)^3 (48/72)^7\approx 0.0176$. Thus the probability for obtaining the same or a better forecast result by random guessing is $p\approx0.022$, well below the significance level 0.05.
Similarly, the probability that in the whole hindcasting and forecasting period, between 1982 and 2021, random guessing would yield an equal or better forecasting performance than our algorithm is $p\approx 4.6\cdotp10^{-5}$.
 
In 2021 $S(t)$ did not cross the threshold from below throughout the year (see Fig. 3), thus predicting the absence of an El~Ni\~no in 2022 with 90\% probability. 
 
\section{System Sample Entropy-based forecast}

\subsection{SysSampEn} 
The SysSampEn was introduced in \cite{Meng2019} as an analysis tool to quantify the complexity (disorder) in a complex system, in particular, in the temperature anomaly time series in the Ni\~no3.4 region. 
In the description and discussion of the method we follow closely \cite{Ludescher2021a}.
The SysSampEn is a generalization of sample entropy (SampEn) and Cross-SampEn \cite{Richman2000}.
SampEn was introduced as a modification of approximate entropy \cite{Pincus1995, Pincus1995b}. It measures the complexity related to the Kolmogorov entropy \cite{Kolmogorov1958}, the rate of information production, of a process represented by single time series. The Cross-SampEn was introduced
to measure the degree of asynchrony or dissimilarity between 2 related time series \cite{Richman2000,Pincus1996}. 
Both have been widely used in physiological fields, however, a complex system such as the climate system is usually composed of several related time series (e.g., curves in Fig. 4). 
Therefore, the SysSampEn \cite{Meng2019} was introduced as a measure of the system complexity, to quantify simultaneously the mean temporal disorder degree of all of the time series in a
complex system and the asynchrony among them. Specifically, it is approximately equal to the negative natural logarithm of the conditional probability that 2 subsequences similar (within a certain tolerance range) for m consecutive data points remain similar for the next p points, where the subsequences can originate from either the same or different time series (e.g., black
curves in Fig. 4), that is,
\begin{equation}
SysSampEn(m, p, l_{eff}, \gamma) = −log(\frac{A}{B}),
\end{equation}
where A is the number of pairs of similar subsequences of length $m + p$, $B$ is the number of pairs of similar subsequences of length
$m$, $l_{eff} \leq l$ is the number of data points used in the calculation for
each time series of length $l$ , and $\gamma$ is a constant which determines
the tolerance range. The detailed definition of SysSampEn for an arbitrary complex system composed of $N$ time series is described in detail in \cite{Meng2019}. When $N = 1$, $p = 1$, and $l_{eff} = l$, the definition is equivalent to the classical SampEn \cite{Richman2000}. 

\begin{figure}[]
\begin{center}
\includegraphics[width=10.5cm]{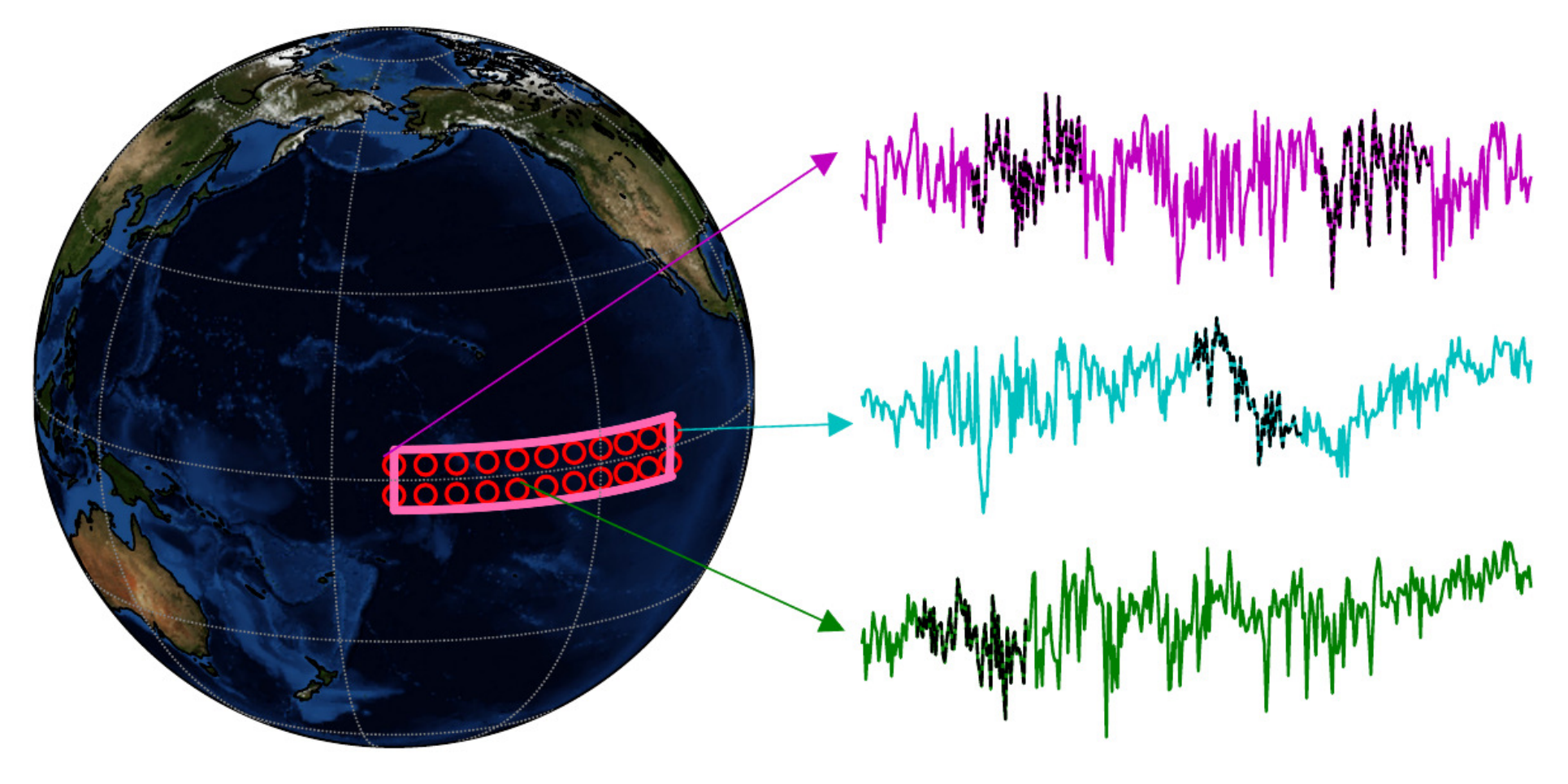}
\caption{The Ni\~no3.4 area and the SysSampEn input data. The red circles indicate the 22 nodes in the Ni\~no 3.4 region with a spatial resolution of $5° \times 5°$ . The curves are examples of
the temperature anomaly time series for 3 nodes in the Ni\~no 3.4 region for
one specific year, and several examples of their subsequences are marked
in black. Figure from \cite{Meng2019}.}
\label{fig4}
\end{center}
\end{figure}

As is the case for SampEn and Cross-SampEn, before the SysSampEn can be
used as an effective tool, appropriate parameter values have to
be identified since only certain value combinations can be used
to estimate a system’s complexity with considerable accuracy. 
The method how to choose parameter values, which yield to a high accuracy when estimating a system's complexity, is described in detail in \cite{Meng2019}. We like to note that identifying the parameters, which yield to a high accuracy, is fully independent of any El~Ni\~no magnitude analysis or forecasts.    
In \cite{Meng2019}, it was found the previous year's ($y-1$) SysSampEn exhibits a strong positive correlation ($r=0.90$ on average) with the magnitude of an El Ni\~no in year $y$ when parameter combinations are used that are able to quantify a system's complexity with high accuracy. 

The linear relationship between SysSampEn and El Ni\~no magnitude enables the prediction of the magnitude of an upcoming El Ni\~no when the current ($y-1$) SysSampEn is inserted into the linear regression equation between the two quantities. If the result is below $0.5\textdegree$C, the absence of an El Ni\~no is forecasted. In other words, SysSampEn values below a certain threshold forecast the absence of an El Ni\~no onset in the following year. Also, an El~Ni\~no onset for the following year is only forecasted if the ONI in December of the current year is below 0.5°C and the SysSampEn is above a certain threshold.

\subsection{Forecast for 2022}

Here we use as input data the daily near-surface (1000 hPa) air temperatures of the ERA5 reanalysis from the European Centre for Medium-Range Weather Forecasts (ECMWF) \cite{ERA5} analysed at a $5\textdegree$ resolution. The last month (Dec) in 2021 is from the initial data release ERA5T, which in contrast to ERA5, only lags a few days behind real-time.
We preprocess the daily time series by subtracting the corresponding climatological mean and then dividing by the climatological standard deviation. We start in 1984 and use the previous years to calculate the first anomalies. For the calculation of the climatological mean and standard deviation, only past data up to the year of the prediction are used. For simplicity, leap days are excluded. 
We apply for the ERA5 data the same parameter combination for the SysSampEn as in \cite{Meng2019}: $m = 30$, $p = 30$, $\gamma=8$ and $l_{eff} = 330$.

Figure 5 shows the results of the analysis. The magnitude forecast is shown as the height of rectangles in the year when the forecast is made, i.e., 1 year ahead of a potential El~Ni\~no onset. The forecast is obtained by inserting the regarded calendar year's SysSampEn value into the linear regression function between SysSampEn and El~Ni\~no magnitude. The regression function for the 2022 forecast is obtained from the best linear fit between the two quantifies for all correctly hindcasted El~Ni\~no events before 2021. 
The red curve shows the ONI and the red shades indicate the El~Ni\~no periods.
The blue rectangles show the correct prediction of an El~Ni\~no in the following calendar year and grey rectangles with a violet border show false alarms. 
There are 11 occurrences of low SysSampEn accompanied by a lower than $0.5°$C ONI in December. In 10 out of these 11 cases, the hindcast was correct. 
White dashed rectangles show correct forecasts for the absence of an El~Ni\~no. Only one El Ni\~no event was not predicted, 2009/10, shown as a pink rectangle. 
There are 13 cases where the forecasted magnitude is above $0.5°$C while the ONI in December is below $0.5°$C, 9 of these cases were followed by an El~Ni\~no, while the remaining 4 cases are false alarms. 
The forecasted El~Ni\~no magnitude for 2022 is $1.90\pm0.23°$C, that is, well above $0.5°$C, as shown by the green rectangle with a black border. The SysSampEn value for 2021 is 1.95, i.e., well above the threshold value of $1.35$. Therefore the method predicts with $69\%$ probability the onset of an El~Ni\~no in 2022.

\begin{figure}[]
\begin{center}
\includegraphics[width=15cm]{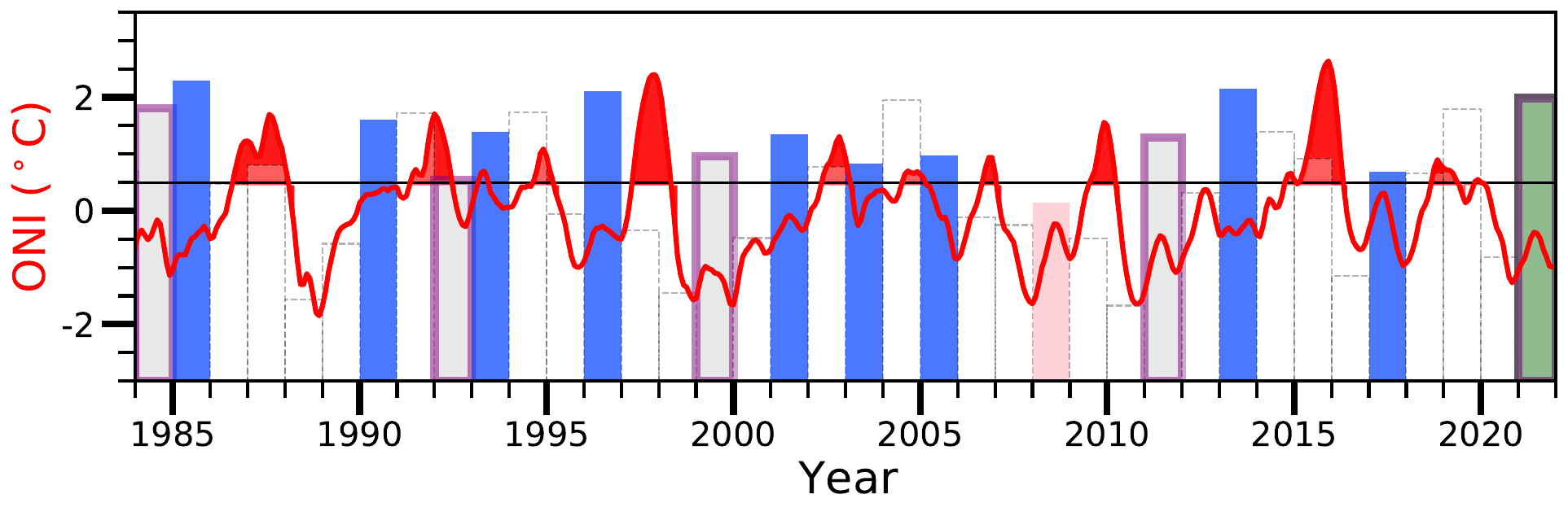}
\caption{Forecasted and observed El~Ni\~no magnitudes. The magnitude forecast is shown as the height of rectangles in the year when the forecast is made, i.e., one year ahead of a potential El~Ni\~no. The forecast is obtained by inserting the regarded calendar year's SysSampEn value into the linear regression function between SysSampEn and El~Ni\~no magnitude. To forecast the following year's condition, we use the ERA5 daily near-surface (1000 hPa) temperatures with the set of SysSampEn parameters ($m = 30$, $p = 30$, $\gamma=8$ and $l_{eff} = 330$), which were obtained in \cite{Meng2019}. 
The red curve shows the ONI and the red shades highlight the El~Ni\~no periods. 
The blue rectangles show the correct prediction of an El~Ni\~no in the following calendar year. 
The onset of an El~Ni\~no in the following year is predicted if the forecasted magnitude is above $0.5°$C and the current year's December ONI is $<0.5°$C. 
White dashed rectangles show correct forecasts for the absence of an El~Ni\~no. 
Grey bars with a violet border show false alarms. The pink rectangle shows the missed El~Ni\~no event 2009/10. 
For 2021 the SysSampEn has a high value of 1.95. When inserting it into the regression function between SysSampEn and El~Ni\~no magnitude, this yield to a forecasted El~Ni\~no magnitude of $1.90\pm0.23°$C. 
There were 13 occurrences of high SysSampEn accompanied by a low ONI in December, as is the case in 2021 (green rectangle). In 9 out of these 13 cases, the hindcast was correct. Thus the method predicts with $69\%$ probability the onset of an El~Ni\~no in 2022.
}
\label{fig5}
\end{center}
\end{figure}

\newpage

\section*{Acknowledgements}
J.L. thanks the East Africa Peru India Climate Capacities (EPICC) project, which is part of the
International Climate Initiative (IKI). The Federal Ministry for the Environment, Nature Conservation and Nuclear Safety (BMU) supports this initiative on the basis of a decision adopted by the German Bundestag.

\end{document}